\documentclass[usenatbib]{mnras}

\usepackage{mathptmx}

\usepackage[T1]{fontenc}
\usepackage{ae,aecompl}

\usepackage{graphicx}	
\usepackage{amsmath}	
\usepackage{amssymb}	

\def \KW  {Konus-\textit{Wind}}

\title[The first observation of an intermediate flare from SGR~1935+2154]{The first observation of an intermediate flare from SGR\,1935+2154}

\author[A. V. Kozlova et al.]{
A.\,V.\,Kozlova$^{1}$\thanks{E-mail: ann\_kozlova@mail.ioffe.ru},
G.\,L.\,Israel$^{2}$,
D.\,S.\,Svinkin$^{1}$,
D.\,D.\,Frederiks$^{1}$,
V.\,D.\,Pal'shin$^{1}$,
\newauthor A.\,E.\,Tsvetkova$^{1}$,
K.\,Hurley$^{3}$,
J.\,Goldsten$^{4}$,
D.\,V.\,Golovin$^{5}$,
I.\,G.\,Mitrofanov$^{5}$ and
\newauthor X.-L.\,Zhang$^{6}$
\\
$^{1}$Ioffe Institute, Politekhnicheskaya 26, St.~Petersburg 194021, Russia\\
$^{2}$INAF - Astronomical Observatory of Rome, Monte Porzio Catone (RM), I-00078, Italy\\
$^{3}$Space Sciences Laboratory, University of California, 7 Gauss Way, Berkeley, CA 94720-7450, USA\\
$^{4}$Applied Physics Laboratory, Johns Hopkins University, Laurel, MD 20723, U.S.A.\\
$^{5}$Space Research Institute, 84/32, Profsoyuznaya, Moscow 117997, Russia\\
$^{6}$Max-Planck-Institut f\"{u}r extraterrestrische Physik, Giessenbachstrasse 1, D-85748 Garching, Germany
}

\date{Accepted 2016 May 8; Received 2016 May 8; in original form 2015 December 23}

\pubyear{}

\begin{document}
\label{firstpage}
\pagerange{\pageref{firstpage}--\pageref{lastpage}}
\maketitle

\begin{abstract}
We report on the bright burst detected by four Interplanetary network (IPN) spacecraft on 2015 April~12.
The IPN localization of the source is consistent with the position of the recently discovered soft gamma-repeater
SGR~1935+2154. From the \KW\,(KW) observation, we derive temporal and spectral parameters of the emission, and
the burst energetics. The rather long duration of the burst ($\sim$1.7~s) and the large measured energy fluence
($\sim2.5\times10^{-5}$~erg~cm$^{-2}$) put it in the class of rare ``intermediate'' SGR flares, and this is the
first one observed from SGR~1935+2154. A search for quasi-periodic oscillations in the KW light curve yields no
statistically significant signal. Of four spectral models tested, optically thin thermal bremsstrahlung and
a single blackbody (BB) function can be rejected on statistical grounds; two more complex models, a cutoff power
law (CPL) and a sum of two BB functions (2BB), fit the burst spectra well and neither of them may be ruled out by
the KW observation. The CPL and 2BB model parameters we report for this bright flare are typical of SGRs;
they are also consistent with those obtained from observations of much weaker and shorter SGR~1935+2154 bursts
with other instruments. From the distribution of double blackbody spectral fit parameters we estimate the
SGR~1935+2154 distance to be $<$10.0~kpc, in agreement with that of the Galactic supernova remnant G57.2+0.8 at 9.1~kpc.
\end{abstract}

\begin{keywords}
pulsars: individual: SGR~1935+2154 -- stars: magnetars -- gamma-rays: stars
\end{keywords}



\section{Introduction}

The history of observations of Soft Gamma Repeaters (SGRs) began over 30 years ago.
The sources were discovered through the detection of recurrent short ($\sim 0.1$\,s) bursts
of hard X-rays/soft $\gamma$-rays by the Konus instrument aboard the Venera~11-14 spacecraft.
The first bursts were initially classified as a subtype of gamma-ray bursts (GRB), one with
a short duration and a soft spectrum~\citep{Mazets1981}.
Now the SGRs are believed to be one of two observational manifestations of ``magnetars'',
isolated neutron stars (NSs) in which the dominant source of free energy is their intense magnetic
field ($B\sim10^{14}-10^{15}$~G)
(see \citealt{Mereghetti2013} for a detailed review and \citealt{McGill} for a recent catalog).

Today we know that the numerous short bursts, with a total energy release
$E_{\mathrm{tot}}$ of~$\sim 10^{38}$--$10^{40}$~erg, are the most common manifestation of
the SGR bursting activity, but there are two other, rarer types of bursts emitted by SGRs:
giant and intermediate flares. The extraordinary giant flares (GFs) are the most intense
Galactic events \citep{Hurley2005}. So far, only three GFs have been observed from three out of 23 confirmed magnetars.
They display an extremely bright, short, hard initial pulse having a huge $E_{\mathrm{tot}} \sim 10^{44}$--$10^{46}$\,erg
followed by a long-duration decaying tail modulated with the NS rotation period.
Quasi-periodic oscillations (QPOs) discovered in the GFs of SGR~1806-20 and SGR~1900+14 \citep{Israel2005,Strohmayer2005}
and, more recently in short bursts (\citealt{Huppenkothen2014a,Huppenkothen2014b}) are expected to help us study the
properties of matter in NSs. Through detection of the frequencies of NS oscillations, it might be possible to deduce neutron
star masses and radii, the equation of state, and other physical properties (e.g., \citealt{Doneva2013, Kokkotas1996}).
The high-fluence intermediate flares (IFs, \citealt{Olive2004}) last from a few seconds up to a few tens of seconds; the brightest
of them have $E_{\mathrm{tot}} \geq 10^{42}$~erg and the peak luminosity reaches $L_{\mathrm{peak}} \geq 10^{43}$~erg~s$^{-1}$~\citep{Mazets1999,Feroci2003,Mereghetti2009,Gogus2011},
less than that of GFs, though orders of magnitude larger than that of the bulk of the SGR burst population.
Although the first observed IFs date back to the early 1980's \citep{Golenetskii1984,Aptekar2001},
only a few dozen flares have been detected so far and the properties of such outstanding events are of special interest.

The source discussed here, SGR~1935+2154, was discovered on 2014 July~5 through a series of
three short bursts \citep{Stamatikos2014,Lien2014}, detected by the \textit{Swift}/Burst Alert Telescope (BAT).
Several days later, its persistent pulsating X-ray counterpart was discovered by the \textit{Chandra} X-ray
Observatory \citep{Atel6370} with a pulse period of~$\sim$~3.2 s. The location of the SGR
as determined by the \textit{Swift}/X-ray Telescope (XRT) \citep{Atel6294} lies very close to the geometric centre
of the Galactic supernova remnant (SNR) G57.2+0.8 \citep{GCN16533}, whose distance is estimated to be
9.1 kpc \citep{Pavlovic2013}. In 2015 February the \textit{Fermi}/Gamma-ray Burst Monitor (GBM) and
BAT observed weak burst activity from the source~--~the first one since the discovery.

In this paper, we report on the first observed IF from SGR~1935+2154, which was detected by
four Interplanetary network (IPN) spacecraft on 2015 April~12.
In Section~\ref{sec:IPN}, we describe the IPN triangulation and localization of the burst.
In Section~\ref{sec:Analysis}, we present the results of temporal and spectral analyses
of the \KW\,(KW) data.
We describe the search for QPOs in the KW light curve in Section~\ref{sec:QPOs}.
Finally, we compare our results to those obtained previously for short bursts and IFs from known SGRs and
discuss the SGR~1935+2154 distance estimate based on the distribution of double blackbody spectral
fit parameters. Unless otherwise specified, all errors refer to 1$\sigma$ confidence limits.

\section{IPN localization}
\label{sec:IPN}
On 2015 April~12, a bright, SGR-like burst was detected by four IPN experiments
-- \emph{INTEGRAL} SPI-ACS \citep{Rau2005}, in a highly elliptical orbit,
\KW\,\citep{Aptekar1995}, in orbit around the Lagrangian point L1,
\emph{MESSENGER} GRNS \citep{Gold2001}, in orbit around Mercury,
and \emph{Mars-Odyssey} HEND \citep{Hurley2006},
in orbit around Mars -- at 0.28, 5.6, 659.6, and 1200.1 light-seconds from Earth, respectively.

\begin{figure}
	\includegraphics[width=\columnwidth]{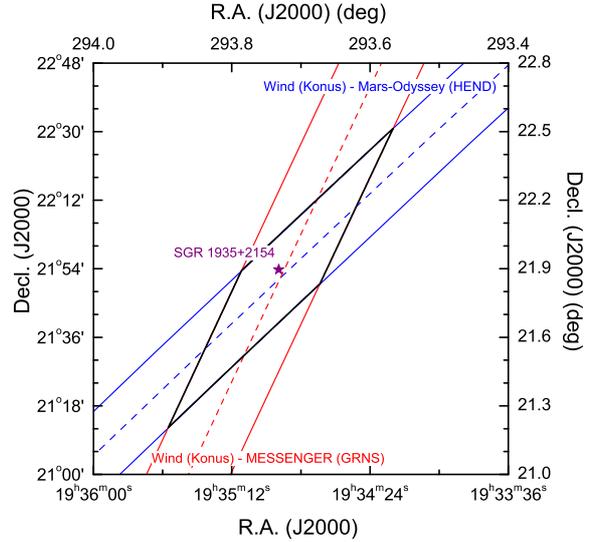}
   \caption{Final, 280~sq.~arcmin IPN error box of 2015 April~12 burst, defined by the $7\farcm08$ wide \emph{Wind}-\emph{Odyssey} and $6\farcm72$ wide \emph{Wind}-\emph{MESSENGER} annuli.
   The position of SGR~1935+2154 is indicated by the star.}
   \label{fig:IPN}
\end{figure}

An initial 1118~sq.~arcmin IPN error box \citep{GCN17699} was derived using KW, \emph{INTEGRAL}, and \emph{Odyssey}.
Later, with \emph{MESSENGER} added, the source was triangulated to a smaller 280~sq.~arcmin IPN error box which is inside the initial box.
The final box is centered at R.A.(J2000) = 19$^\rmn{h}$34$^\rmn{m}$55$^\rmn{s}$, decl.(J2000) = $+21\degr$51$\arcmin$49$\arcsec$ and its maximum dimension is $1\fdg34$ (the minimum one is $6\farcm72$).
The position of SGR~1935+2154 \citep{Atel6294} lies inside the final error box, $1\farcm97$ from its centre (Fig.~\ref{fig:IPN}).

\section{Konus-\emph{WIND} observation}
\label{sec:Analysis}
The burst triggered KW at $T_{0}=41064.683$~s UT (11:24:24.683) on 2015 April~12.
The propagation delay from \emph{Wind} to Earth is 1.361~s for this burst;
correcting for this factor, the KW trigger time corresponds
to the Earth-crossing time 41066.044~s UT (11:24:26.044).

\subsection{Time history}

\begin{figure}
	\includegraphics[width=\columnwidth]{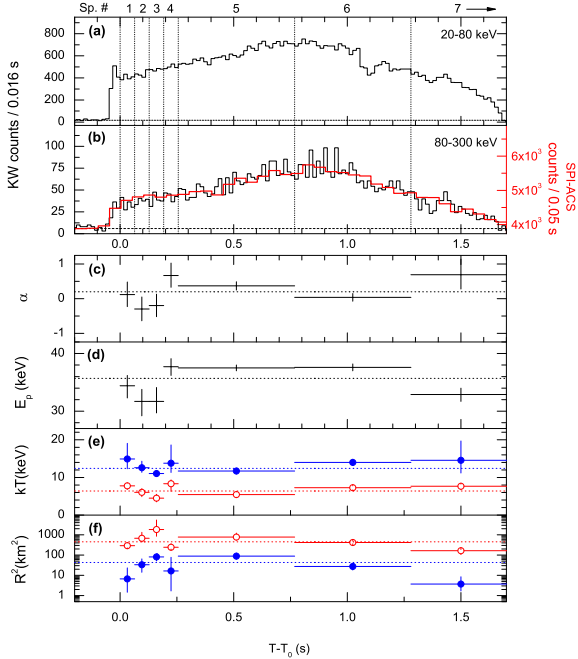}
   \caption{Light curves of the burst recorded by Konus-\textit{Wind} in the G1 (20--80 keV, panel (a)) and G2 (80--300 keV; panel (b), black line)
   energy ranges with 16~ms resolution; the \emph{INTEGRAL} SPI-ACS light curve ($\gtrsim$80 keV, 50~ms resolution) is shown with a red line in panel (b).
   The vertical dotted lines denote the intervals over which the KW spectra were accumulated; the right boundary of spectrum~7 ($T_0+9.472$~s) is not shown.
   The KW count rates are dead-time corrected and the horizontal dashed lines indicate the background levels.
   The \emph{INTEGRAL} time scale is corrected for the burst propagation between the spacecraft.
   Time-resolved CPL and 2BB fit parameters are given in panels (c) to (f); low- and high-kT components of 2BB are shown by open and filled symbols, respectively;
   the horizontal dotted lines indicate the value of the parameters measured for the time-integrated spectrum.}
   \label{fig:lightcurve}
\end{figure}

The event time histories were
recorded in three energy bands: G1 (20--80 keV), G2 (80--300 keV) and G3 (300--1200 keV)
with a time resolution of 2~ms from ${T_{0}-0.512}$~s to $T_{0}+0.512$~s and 16~ms afterwards.
The burst light curves in bands G1 and G2 (Fig.~\ref{fig:lightcurve}) show a single, very bright pulse which starts at~$\sim T_0-0.062$~s with a sharp ($<$10 ms)
rise, peaks at $\sim T_0+0.800$~s and decays to background level at $\sim T_0+1.680$~s. KW observed no statistically significant
emission above $\sim$300~keV (the G3 band), which, accounting for the burst brightness, suggests a soft energy spectrum.
The total burst duration $T_{100}=1.742$~s was determined at the $5 \sigma$
level in the G1+G2 energy band (20--300 keV). The corresponding values of $T_{90}$ and $T_{50}$
are $1.412\pm0.016$ and $0.654\pm0.016$~s, respectively.

While the KW count rate in this burst reaches $\sim 5\times10^4$~counts~s$^{-1}$,
it does not exceed $\sim$15\% of the saturation level, which makes the standard KW dead time (DT) correction procedure
reliable (the procedure uses a non-paralyzable correction with a DT of 2--5 microseconds, depending on the energy band;
details of the KW DT corrections at high count rates can be found in \citealt{Mazets1999b}).
Given the soft energy spectrum, the DT-corrected KW light curve in the G2 band (80--300 keV) can be directly compared to the SPI-ACS data (E$\gtrsim$80~keV).
At the SPI-ACS count rate of $\sim 1.1\times10^5$ counts~s$^{-1}$ observed in this burst the DT effects are negligible \citep{Mereghetti2005};
and a clear similarity in the shapes of the two instrument light curves (panel (b) in Fig.~\ref{fig:lightcurve})
demonstrates the correctness of our analysis.

\subsection{Spectral analysis and the burst energetics}

During the burst, the instrument measured seven multichannel energy spectra covering a wide energy
range from 20~keV to 14~MeV; the accumulation times are given in Tab.~\ref{tab:spec}.
For the spectral fits, we use only the 20 to 250~keV energy interval since no emission was detected at higher energies.
At high count rates a pile-up effect in the analog electronics can distort
the low-energy part of the KW instrumental spectra (e.g., \citealt{Frederiks2013}).
Our previous simulations show that for events with soft, SGR-like incident spectra and count rates similar to those in this event,
the scale of the resulting distortion does not exceed statistical errors in the corresponding spectral channels.
Accordingly, spectral fits we made to the simulated data did not reveal a statistically significant difference
between the model and the best-fit spectral parameters. Thus our analysis of the SGR~1935+2154 burst spectra
relies on standard fitting procedures with no special precautions due to high count rates being taken.

The spectral analysis was performed in XSPEC, version 12.8 \citep{Arnaud1996}, by applying two spectral models,
which have been shown to be the best-fits to the broadband spectra of SGR bursts \citep[e.g.][]{Feroci2004, Olive2004, Lin2012, Horst2012}.
The first one is a sum of two blackbody functions with the normalization proportional to the surface area (2BB).
The second model is a power law with an exponential cutoff (CPL), parametrized as $E_{\mathrm{p}}$: $f(E)\propto E^{\alpha} \exp(-(2+\alpha)E/E_{{\mathrm{p}}})$,
where $\alpha$ is the power-law photon index and $E_{\mathrm{{p}}}$ is the peak energy in the $\nu$F$_\nu$ spectrum.
We also tried to fit the spectra to a single blackbody (BB) function
and to an optically thin thermal bremsstrahlung (OTTB, $f(E)\propto E^{-1}\exp(-E/kT_{\mathrm{OTTB}})$) and found that both models may be rejected on statistical grounds.

A summary of the KW spectral fits with 2BB and CPL models is presented in  Tab.~\ref{tab:spec}.
Two methods were used to obtain the best-fit parameters for any given spectral model.
In the first method, the raw count rate spectra were rebinned in order to have at least 10~counts per energy bin,
and fitted using $\chi^2$ minimization. The alternative method uses the Castor C-statistic (C-stat) minimization
and the spectra rebinned to have a minimum of one count per bin.
For spectra 1-4, with the short accumulation times and poor count statistics in higher energy channels,
the fits using C-stat gave smaller uncertainties in the parameters than those using the $\chi^2$ statistic, but they are consistent
with each other within the uncertainties. In these cases we report the results obtained with the C-statistic
and provide a quality of the corresponding $\chi^2$ fit for reference. Otherwise, the results obtained with $\chi^2$ are provided.
We note that spectrum~7 was measured from $T_0$+1.280~s to $T_0$+9.472~s and no burst emission was detected after $T_0+1.680$~s.
For this spectrum, BB radii and the corresponding luminosities obtained from the XSPEC fits were re-calculated using
the accumulation interval 1.280--1.680~s; accordingly, for the time-integrated spectrum
the BB normalizations are given for the interval 0.0--1.680~s.

\begin{figure}
	\includegraphics[width=\columnwidth]{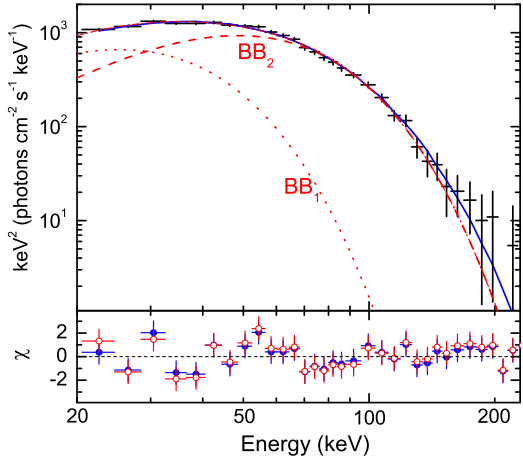}
   \caption{Time-integrated $\nu$$F_\nu$ spectrum of the flare: KW data (symbols); the CPL model (blue solid line);
   the 2BB model (red dashed-dotted line) and its low-$kT$ and high-$kT$ components (red dotted and dashed lines, respectively).
   The bottom panel shows the fit residuals: filled and open symbols represent CPL and 2BB, respectively.}
   \label{fig:specint}
\end{figure}

Both CPL and 2BB models fit the time-integrated (TI) spectrum (Fig.\ref{fig:specint}) and all seven time-resolved (TR) spectra well,
with $\chi^2$/dof=1.23 (30 dof) in the worst case and a null hypothesis probability of $>$0.18 for all fits.
When comparing the fit statistic for individual spectra, the difference
between CPL and 2BB fits, $\Delta\chi^2_\mathrm{CPL-2BB}$, lies between $-5.4$ and $+2.7$ with nearly equal numbers
of positive and negative values of $\Delta\chi^2_\mathrm{CPL-2BB}$ obtained
(the behaviour of C-stat is very similar and, hereafter, we focus on the $\chi^2$ statistic only).

Since these two models are not nested, the most preferred model can not be chosen
based on a test statistic (TS) with known reference distribution (e.g., using an $F$-test or a likelihood ratio test).
This difficulty can be overcome e.g. by employing a Bayesian approach when a posterior predictive distribution of
an arbitrary TS is created via Monte Carlo simulations of the model parameters.
Following the recipe from \cite{Protassov2002}, we ran a set of simulations in XSPEC in order to evaluate the capability of
KW to distinguish between the two models for the SGR~1935+2154 spectra.
For each of the time intervals we simulated two sets of 1000 spectra using CPL or 2BB as a null model and fitted the simulated spectra with both models.
From the fits, we built posterior predictive distributions of $\Delta\chi^2$ and calculated the posterior predictive $p$-values.
The latter represent, in our case, the probability for an incident CPL (2BB) spectrum to be fit by 2BB (CPL) model
with more extreme $\Delta\chi^2$ than that obtained from the real spectrum.
For none of the time intervals did we find $p<10^{-2}$.
CPL can be preferred to 2BB ($p \lesssim 0.05$) only for spectrum \#6 ($p=0.051$) and for the TI spectrum ($p=0.019$),
while 2BB cannot be favoured at a level better than $p=0.173$ achieved for spectrum \#1.
Proceeding from this, and from the good agreement found for both spectral functions with the measured data,
we conclude that neither CPL nor 2BB may be ruled out by the KW observations and consider below the results obtained
with both models.

The TI spectrum 1-7 is well fitted by the CPL model ($\chi^2=32.4/31$\,dof) with
$\alpha$ of $\simeq$0.20 and $E_{\mathrm{p}}\simeq$35.7~keV.
The time-resolved spectral parameters are plotted in panels (c--f) in Fig.~\ref{fig:lightcurve}.
The TR peak energies of the CPL model in individual spectra range from $\sim$31.7 to $\sim$37.5~keV,
with a slight correlation to the energy flux.
The photon index shows more prominent variation: $\alpha$ changes over a wide range of hard values,
between -0.3 and +0.7, showing no apparent dependence on the emission intensity.

The 2BB model applied to the TI spectrum also yields a reasonably good fit ($\chi^2=37.0/30$\,dof).
The temperatures of the soft and hard BB components are ${kT_1\simeq 6.4}$ and ${kT_2\simeq 12.4}$ keV, and the
corresponding radii of the emitting areas (calculated at a distance of 10~kpc) are $R_1\simeq$~21.3
and $R_2\simeq$~6.6~km, respectively.

The TR $kT_1$ values vary in the $\sim$4.5--8.5~keV range.
Although $kT_2$ has large uncertainties in some spectra, the best-fit values are located between 11 and 15~keV
and show smaller relative fluctuations than $kT_1$, with $\sigma(\mathrm{log}kT_2)$=0.050 as compared to $\sigma(\mathrm{log}kT_1)$=0.098.
Finally, the TR BB radii as well as the derived BB luminosities generally follow the count rate evolution.
A simple statistical test suggests that the fluctuations observed in the CPL and 2BB model parameters
can hardly be interpreted as purely statistical; fits with a constant level lead to $\chi^2>$11.2/6~dof
for all the spectral parameters, indicating that some spectral variability is present over the duration of the burst.

\begin{figure}
   \centering
  \includegraphics[width=\columnwidth]{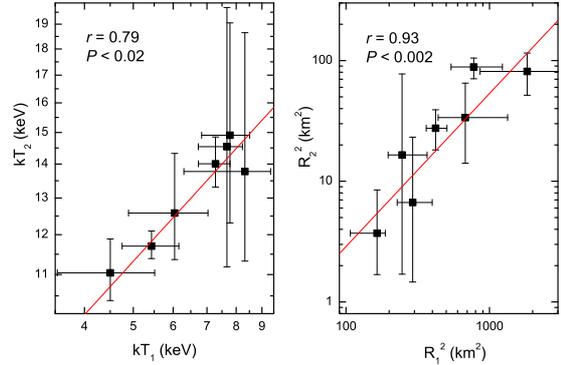}
   \caption{Correlations between the cool and hot blackbody parameters: temperatures (left panel) and emission areas (right panel).
   Spearman-rank correlation coefficients $r$ and chance probabilities $P$ are indicated in each panel.
   Solid lines represent the best-fit power-law approximations (see text).}
   \label{fig:kT2vskT1}
\end{figure}

Fig.~\ref{fig:kT2vskT1} shows a correlation between the cool and hot BB temperatures derived for individual spectra (left panel),
as well as between the time-resolved BB emission areas (right panel).
Spearman-rank correlation coefficients $r$ and chance probabilities $P$ are indicated in each panel.
While the emission area correlation, with $r$=0.93, is almost significant at the >3$\sigma$ level,
the temperatures are less strongly correlated. When fitted by a power law, the slopes of the correlations are
0.52$\pm$0.14 and 1.27$\pm$0.17 for the temperatures and the radii, respectively.
Finally, the blackbody radii are anti-correlated with corresponding BB temperatures (see Fig.~\ref{fig:R2vskT} and discussion below).

From the CPL spectral fits, we estimate the total energy fluence of the burst $S$ to be
$(2.50\pm 0.03)\times 10^{-5}$\, erg cm$^{-2}$ and the peak energy flux $F_{\mathrm{max}}$ to be
$(2.15\pm 0.13)\times 10^{-5}$\, erg cm$^{-2}$ s$^{-1}$ in a 16~ms time interval starting at
$T_0+0.800$\,s; both values are calculated in the 20--200~keV energy range.
From the 2BB fits, the bolometric fluence and peak flux estimates are $\sim3.3\times 10^{-5}$\, erg~cm$^{-2}$ and $\sim3.0\times 10^{-5}$\,erg~cm$^{-2}$~s$^{-1}$, respectively.
The low-$kT$ BB component contributes about 27~per~cent to the total 20--200 keV flux and about 42~per~cent to the total bolometric fluence.

\begin{table*}
\begin{minipage}{170mm}
	\centering
	\caption{Spectral fits with 2BB and CPL models.}
	\label{tab:spec}
    \tiny
	\begin{tabular}{ccccccccccccc} 
		\hline
		& & \multicolumn{7}{c}{2BB Model}  &\multicolumn{3}{c}{CPL Model} \\[2pt]
		Spectrum & Interval  & kT$_1$ & Norm$_1$ & L$_1^a$ & kT$_2$ & Norm$_2$ & L$_2^a$ & $\chi^2/$dof$^{b}$ & $\alpha$ & E$_{\mathrm{peak}}$ & $\chi^2/$dof$^{b}$\\[2pt]
		& (s from $T_0$)& (keV) & R$_{\text{km}}^2/{d_{10}^2}$ & ($10^{39}$~erg)& (keV) & R$_{\text{km}}^2/{d_{10}^2}$ & ($10^{39}$~erg)& & & (keV) &\\
		\hline
1	&0.0 - 0.064   &$7.8^{+0.7}_{-0.9}$	&$292^{+107}_{-64}$	    & $137^{+22}_{-33}$ & $14.9^{+4.1}_{-2.6}$ & $6.7^{+16.6}_{-5.2}$ & $49.0^{+34}_{-23}$ &11.9/25 [4.8/15]		
&$0.12^{+0.36}_{-0.35}$		&$34.4^{+1.8}_{-2.1}$	&14.5/26 [6.9/16]\\[3pt]
2	&0.064 - 0.128 &$6.0^{+1.0}_{-1.1}$	&$678^{+660}_{-239}$    & $116^{+29}_{-27}$ & $12.6^{+1.8}_{-1.2}$  &$33.7^{+31.3}_{-19.6}$  & $108^{+33}_{-37}$ &10.5/21 [8.3/15]		
&$-0.30^{+0.35}_{-0.34}$	&$31.7^{+2.1}_{-2.5}$	 &10.3/22 [8.1/16]\\[3pt]
3	&0.128 - 0.192 &$4.5^{+1.0}_{-1.0}$	&$1830^{+3650}_{-970}$	& $96^{+20}_{-19}$ &  $11.0^{+0.9}_{-0.7}$  &$81.3^{+34.0}_{-26.7}$  & $155^{+20}_{-25}$ &26.7/24 [15.0/15]	
&$-0.20^{+0.33}_{-0.32}$	&$31.7^{+2.4}_{-2.0}$	 &28.3/25 [16.4/16]\\[3pt]
4	&0.192 - 0.256 &$8.5^{+1.0}_{-2.0}$	&$246^{+122}_{-49}$	& $151^{+50}_{-86}$ &$13.8^{+4.9}_{-2.5}$ &$16.5^{+60.9}_{-14.8}$  & $76^{+89}_{-52}$ &13.7/21 [8.4/15]		
&$0.67^{+0.36}_{-0.34}$		&$37.7^{+1.4}_{-1.5}$	&14.2/22 [8.8/16]\\[3pt]
5	&0.256 - 0.768 &$5.4^{+0.7}_{-0.7}$	&$776^{+454}_{-236}$   & $87^{+16}_{-11}$ &  $11.7^{+0.4}_{-0.3}$   &$88.4^{+16.5}_{-17.8}$	& $213^{+15}_{-20}$ &26.2/28	
&$0.36^{+0.12}_{-0.12}$		&$37.6^{+0.5}_{-0.5}$	 &25.0/29\\[3pt]
6	&0.768 - 1.280 &$7.3^{+0.5}_{-0.5}$	&$421^{+84}_{-58}$    & $152^{+21}_{-21}$ & $14.0^{+0.8}_{-0.7}$  &$27.5^{+11.6}_{-9.3}$	& $136^{+23}_{-23}$ &37.0/30	
&$0.04^{+0.11}_{-0.11}$		&$37.6^{+0.6}_{-0.6}$	 &31.6/31\\[3pt]
7	&1.280 - 9.472$^{c}$&$7.7^{+0.6}_{-1.0}$ & $165^{+23}_{-60}$& $73^{+10}_{-25}$ &  $14.5^{+5.2}_{-3.4}$   &$3.7^{+7.8}_{-2.0}$	& $20^{+26}_{-11}$ &23.1/30	
&$0.70^{+0.46}_{-0.43}$		&$32.9^{+1.1}_{-1.2}$	 &25.8/31\\[3pt]
1 - 7$^{d}$	&0.0 - 9.472$^{c}$&$6.4^{+0.4}_{-0.4}$&$455^{+73}_{-55}$ & $95^{+12}_{-11}$ &$12.4^{+0.4}_{-0.4}$&$43.5^{+8.8}_{-8.1}$  & $136^{+12}_{-14}$ &37.0/30	
&$0.20^{+0.08}_{-0.08}$		&$35.7^{+0.3}_{-0.3}$	 &32.4/31\\[3pt]
		\hline
	\end{tabular}
    \normalsize
	\end{minipage}
\flushleft
    \tiny
$^{a}$ The luminosity of blackbody components calculated at $d$=10~kpc.\\
$^{b}$ C-stat/dof for spectra 1 to 4; the quality of the corresponding $\chi^2$ fit is given in square parentheses for reference. \\
$^{c}$ The 2BB model radii and luminosities for spectra 7 and 1-7 are calculated using intervals 1.280--1.680~s and 0.0--1.680~s, respectively (see text).\\
$^{d}$ The time-integrated spectrum.
    \normalsize
\end{table*}

\section{Search for QPOs}
\label{sec:QPOs}
\begin{figure}
   \centering
   \includegraphics[width=0.7\columnwidth]{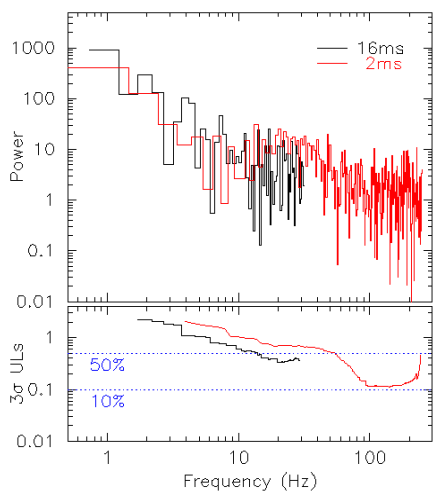}
   \caption{Power spectrum produced from the KW light curve in the G1+G2 energy range (upper panel).
   Curves representing the upper limits to the non-detection of pulsations are in the lower panel. The two
   horizontal dashed lines represent the 10 per cent and 50 per cent upper limits to the pulsed fraction.
   Lines of different colours correspond to the different time resolutions. }
   \label{fig:qpo}
\end{figure}

We searched for pulsations in the 2~ms and 16~ms G1+G2 light curves by using the Fast Fourier Transform method without finding any
statistically significant signal. Following the recipe described in \cite{Israel1996} we derived upper limits to the pulsed
fraction at a 3$\sigma$ confidence level of~$>$50 per cent for frequencies in the 5--60 Hz range and 10--30 per cent between 60 and 250~Hz
(see Fig.~\ref{fig:qpo}).

When compared to the QPO amplitudes detected in tails of GFs these limits are not very strict.
The main QPOs in the GFs of SGR~1806-20 and SGR~1900+14 \citep{Israel2005,Strohmayer2005} have pulsed fractions
of $\sim$5 per cent at $\sim$20~Hz, $\sim$50 per cent at $\sim$30~Hz, $\sim$10 per cent at $\sim$150~Hz and $\sim$20 per cent at $\sim$260~Hz.
In all cases our inferred 3$\sigma$ upper limits are well above or slightly above these values.

\section{Discussion}
\label{sec:Discussion}
The rather long duration of the burst (longer than half the rotation period) along with the large measured
energy fluence put it in the class of ``intermediate'' SGR flares, and this is the first one observed from SGR~1935+2154.
The burst profile is rather structureless; it lacks a prominent initial peak seen in some SGR bursts of comparable
(or longer) duration \citep{Mazets1999,Aptekar2001,Olive2004}
and does not show any evidence of an extended, modulated tail observed previously in a number of IFs \citep[e.g.][]{Mereghetti2009,Gogus2011}.
In the Type~A/Type~B morphological classification of magnetar bursts \citep{Woods2005}
such an event may be classified as Type~A.

A spectral model that best describes the SGR~1935+2154 burst spectra cannot be selected unambiguously from our analysis.
Of the four spectral functions tested, only two simple models, OTTB and single BB, can be rejected on statistical grounds
while two more complex models, CPL and 2BB, fit the TI spectrum and all seven TR spectra reasonably well
and neither of them may be ruled out by our observations. The main reason behind this is that, given the observed count statistics
and the derived model parameters, the CPL and 2BB spectra can successfully mimic each other in the relatively hard (20--250~keV)
KW spectral band. Similar results were obtained in the \emph{Fermi}/GBM studies of SGR~J0501+4516 and SGR~J1550-5418 bursts
in the softer 8--200~keV band \citep{Lin2011,Horst2012}.
However, with the use of the \emph{Swift}/XRT 0.5--10~keV data,
it was shown that the broadband (0.5--200~keV) spectra of SGR~J1550-5418 bursts observed simultaneously with GBM and XRT
are better described with two blackbody function than with the Comptonized (CPL) model \citep{Lin2012}.
Thus, broadband studies of SGR~1935+2154 are needed to reach more conclusive results on its spectral behaviour.

The CPL model is intended to approximate the unsaturated Comptonization spectrum,
and its implications in the context of magnetars are presented in detail in \cite{Lin2011}.
The $E_{\mathrm{p}}$ value we obtained from the CPL model fits, $E_p\sim$~30--40~keV,
is typical for SGR bursts (see, e.g., \citealt {Aptekar2009,Lin2011,Feroci2004}).
The OTTB function, which is often considered for SGR spectra above 15~keV \citep[e.g.][]{Aptekar2001},
does not fit the TI spectrum nor any individual burst spectra after $T_0+0.128$~s ($\chi^2$/dof$ > $2).
Accordingly, the hard photon index obtained from the CPL fits ($\alpha \gtrsim -0.3$) is inconsistent with the OTTB slope of $-1$.
Our result for the CPL photon index is close to those reported for SGR~0501+4516 \citep{Aptekar2009,Lin2011} and SGR~1900+14 \citep{Feroci2004},
and is significantly harder than the OTTB-like slope initially derived for SGR~J1550-5418 from the GBM observations by \cite{Horst2012}.
It was shown by \cite{Lin2012} that the broadband XRT+GBM spectra of SGR~J1550-5418 bursts better constrain the photon indices,
which become harder than the ones derived from the GBM-only data.
They also noticed that a classical, unsaturated Comptonization model has difficulty in generating spectral slopes with $\alpha>-1$
and that such flat spectra might naturally be expected to exhibit a more truly thermalized character.
A generalized Comptonization model (CompTT; \citealt{Titarchuk1994}), where soft photons are upscattered in a hot plasma taking into account
relativistic effects, have been shown to be similar to the 2BB performance in the broadband study of SGR~1900+14 bursts with \emph{Swift}/BAT and XRT \citep{Israel2008}.
We fitted this model (implemented as \texttt{CompTT} in XSPEC) to the SGR~1935+2154 spectra with good count statistics
and obtained $\chi^2$ values between those of CPL and 2BB.
This demonstrates the good statistical performance of CompTT on the IF spectra, but a more detailed discussion of this model is beyond the scope of this paper.

\begin{figure}
   \centering
  \includegraphics[width=\columnwidth]{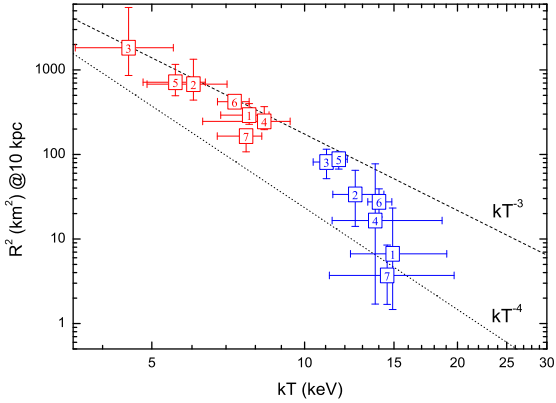}
   \caption{Squares of the radii of the emitting areas as a function of their temperatures. The soft BB component is shown
   in red and the hard BB in blue. Numbers correspond to the spectral intervals.
   We also plot the $R^2\propto kT^{-3}$ and $R^2\propto kT^{-4}$ power laws (the
   latter corresponds to the relation for a pure BB with $L=3\times 10^{40}$~erg~s$^{-1}$).}
   \label{fig:R2vskT}
\end{figure}

The most plausible interpretation of the 2BB model is the emission originating from two hot spots with
different temperatures near or on the neutron star surface or in its magnetosphere where local thermodynamic equilibria are achieved.
For the TI spectrum, the temperatures and radii we obtained for the 2BB model are typical of that for other SGR sources:
$kT_{1}\sim$ 3--7~keV, $R_{1}\sim$ 10--30~km and $kT_{2}\sim$ 10--20~keV, $R_{2}\sim$~4~km;
see \cite{Nakagawa2007} for SGR~1806-20 and SGR~1900+14, \cite{Esposito2008} for SGR~1627-41,
\cite{Horst2012} for SGR~J1550-5418, and \cite{Lin2011} for SGR~J0501+4516. Specifically, the derived
2BB model parameters are in reasonable agreement with the results reported previously for intermediate
bursts: $kT_{1} \sim$ 4.8~keV, $R_{1} \sim$ 30~km and $kT_{2} \sim$ 9.0~keV, $R_{2} \sim$
5.7~km \citep{Israel2008, Olive2004}.
Finally, the spectral parameters we measured for the IF are consistent with
those obtained from observations of much weaker and shorter bursts from SGR~1935+2154 in 2014 July \citep{Lien2014}
and in 2015 February \citep{GCN17496}. This suggests that similar physical processes may be
responsible for the IF and the weak bursts despite orders of magnitude difference in the amount of energy released.

Using the 2BB fits described in Section~\ref{sec:Analysis}, we calculated the soft and hard BB luminosities for each of the seven TR spectra (Tab.~\ref{tab:spec}).
All the derived bolometric luminosities are over 10$^{40}$~erg~s$^{-1}$, and hence there is a slight hint of the saturation effect of the
low-$kT$ BB luminosity as previously noted for SGR~1900+14~\citep{Israel2008}. In order to further investigate this trend, we studied the $R^2$
versus $kT$ distribution (see Fig.~\ref{fig:R2vskT}). The sharp edge in the distribution of the data described by the $R^2\propto kT^{-3}$ relation
indicates the presence of the saturation. So we can use the magnetic Eddington luminosity formula derived in~\cite{Paczynski1992}:
\begin{equation}
L_{\mathrm{Edd,B}}\approx 2\times 10^{40}\left(\frac{B}{B_{\mathrm{QED}}}\right)^{4/3} \left(\frac{R}{R_{\mathrm{NS}}}\right)^{2/3},
\end{equation}
where $B$ is the magnetic field at the radius $R$, $B_{\mathrm{QED}}\simeq 4.4 \times 10^{13}$~G is magnetic field critical value and $R_{\mathrm{NS}}$
is the neutron star radius for which we assume a typical value of 10~km.
It can be rewritten in terms of the distance $d$ and peak flux $F_{\mathrm{max}}$ as

\begin{multline}
\left(\frac{d}{\mathrm{kpc}}\right) \simeq 0.4 \times \left(\frac{F_{\mathrm{max}}}{10^{-5}\,\mathrm{erg}\, \mathrm{cm}^{-2}\,\mathrm{{s}^{-1}}}\right)^{-1/2}
\left(\frac{kT_{\mathrm{break}}}{\mathrm{keV}}\right)^{5/4} \times \\
\times \left(\frac{B_{\mathrm{surf}}}{10^{14}\,\mathrm{G}}\right)^{1/4}
 \left(\frac{R_{\mathrm{NS}}}{10\, \mathrm{km}}\right)^{5/8},
\end{multline}
where $kT_{\mathrm{break}}$ is the energy at which the data in the $R^2$ versus $kT$ distribution
start departing from the relation $kT^{-3}$.
Now we can estimate an approximate value for the source distance by using the saturated flux recorded for
the source and the magnetic field strength inferred by timing analysis. For the surface magnetic field $B_{\mathrm{surf}}$ we use
a dipolar magnetic field value of~$\sim 2.2 \times 10^{14}$~G inferred for SGR~1935+2154 by \cite{Israel2016} from \emph{Chandra} observations.
With the value of $kT_{\mathrm{break}}$ for this burst lying in the 12--15 keV range we derive a distance of 7.4--9.8 kpc.
However, given that we have seen only one bright IF from this source, we cannot be sure that the luminosity in the April 12 burst is close to the maximum observable
from SGR~1935+2154 and a brighter burst would situate the source closer to us.
So, in this work, we estimate the SGR~1935+2154 distance to be $<$10.0~kpc, in agreement with that of the Galactic supernova remnant G57.2+0.8.
Assuming isotropic emission at 9.1~kpc, the total bolometric energy release in the flare is $ \sim 3.3\times 10^{41}$~erg
and the bolometric peak luminosity is $ \sim 3.0\times 10^{41}$~erg~s$^{-1}$; both of them lie close to the lower end of the IF range.

\section*{Acknowledgements}
We would like to thank the anonymous referee for providing comments that helped to improve the paper.
We are grateful to W.~Boynton, C.~Fellows, K.~Harshman, A.~S.~Kozyrev, M.~L.~Litvak, A.~B.~Sanin, A.~Rau and A.~von~Kienlin for the use of their data in the IPN triangulation.
DDF acknowledges support from RFBR grant 15-02-00532a.
KH acknowledges support for MESSENGER data analysis from NASA Grant NNX07AR71G.




\bibliographystyle{mnras}
\bibliography{ms}


%
%
%


\bsp	
\label{lastpage}
\end{document}